\documentclass[11pt,twoside]{article}

\usepackage{asp2004}
\usepackage{epsf}
\usepackage{graphicx}
\usepackage{lscape}

\newcommand{\kms}{\ensuremath{{\rm km\,sec^{-1}}}}                   % km/sec
\newcommand{\msunyr}{\ensuremath{\mathit{M}_{\odot}{\rm yr}^{-1}}}   % msun/yr
\newcommand{\mdot}{\ensuremath{\dot{M}}}                             % mass loss rate
\newcommand{\rstar}{\ensuremath{\mathit{R}_{\star}}}                 % stellar radius
\newcommand{\teff}{\ensuremath{\mathit{T}_{\rm eff}}}                % effectieve temperatuur
\newcommand{\mdu}{\ensuremath{\cdot 10^{-6} \msunyr}}
\newcommand{\Ha}{H$_{\alpha}$}
\newcommand{\fcl}{\ensuremath{f_{\rm cl}}}

\bibpunct{(}{)}{,}{a}{}{,}

\begin{document}
\title{Stellar winds from massive stars - What are the REAL mass-loss rates?}    %%% Fill in title
\author{Joachim Puls$^1$, Nevena Markova$^2$ \& Salvatore Scuderi$^3$}   %%% Fill in author names
\affil{$^1$Universit\"ats-Sternwarte, Scheinerstr. 1, D-81679 M\"unchen, Germany\\
      $^2$Institute of Astronomy, Bulgarian National Astronomical Observatory, 
      P.O. Box 136, 4700 Smoljan, Bulgaria\\
      $^3$INAF - Osservatorio Astrofisico di Catania, Via S. Sofia 78, I-95123
      Catania, Italy}

\begin{abstract}
%%% Abstract to run on from here.
We discuss recent evidence that currently accepted mass-loss rates may need
to be revised downwards, as a consequence of previously neglected
``clumping'' of the wind. New results on the radial stratification of the
corresponding clumping factors are summarized. We investigate the influence
of clumping on the ionization equilibrium of phosphorus, which is of major
relevance when deriving constraints on the clumping factors from an analysis
of the FUV P{\sc v} resonance lines.
\end{abstract}

%%% MAIN BODY OF TEXT GOES HERE. CONSULT "INSTRUCTIONS FOR AUTHORS USING
%%% LATEX2E MARKUP", SECTIONS 2.3-2.6 FOR HELP WITH EQUATIONS, FIGURES,
%%% AND TABLES.
\section{Introduction}

While our understanding of the outflows from luminous OB-stars was thought
to be well established, recent evidence indicates that currently accepted
mass-loss rates {\it may} need to be revised downwards {\it by as much as a
factor of ten}. This is a consequence of previously neglected ``clumping''
of the wind, which affects mostly those diagnostics which are sensitive to
the {\it square} of the density, $\rho$ (such as recombination lines or
free-free continua).

Considering that numerous stellar-evolution calculations have demonstrated
that changing the mass-loss rates of massive stars by even a factor of two
has a dramatic effect on their evolution (e.g., \citealt{Meynet94}), it is
evident that such revisions would have enormous implications, not only
regarding evolution, but also regarding the feed-back from massive stars. In
this article, we will summarize the knowledge which has been accumulated
lately and consider the question concerning the REAL mass-loss rates from
massive star winds.

\section{Standard mass loss diagnostics of luminous OB-stars}
\label{diag}

Traditionally, the mass-loss rates, \mdot, of luminous OB stars have been
inferred from, primarily, three types of measurement (see also de Koter,
this volume):\\ 
\underline{1. The strengths of UV P-Cygni profiles} were the first
diagnostics to be used to measure wind-densities. This approach has been
pioneered by \citet{LM76} in their work on the ``Mass ejection from the O4f
Star Zeta Puppis", where they derived a mass-loss rate of $7.2 \pm 3.2
\mdu$, a number which is still valid! The most commonly used method to
analyze P-Cygni lines is the so-called SEI-method \citep{Lamersetal87},
which has been firstly applied by \citet{GL89}. Though most of the UV
P-Cygni lines are resonance lines from dominant ions being {\it linearly}
dependent on density and thus remaining uncontaminated from {\it direct}
clumping effects (but see Sect.~\ref{pv}), only the product $\mdot q$ can be
inferred from these lines, where $q$ is the ionization fraction of the
corresponding ion. Moreover, most of these lines are inevitably saturated in
stars with strong winds, since they arise from {\it abundant} elements, and
only lower limits on \mdot\ can be provided in these cases. \\
\underline{2. Thermal radio and FIR continuum emission} samples the
outermost and intermediate region of the wind ($\mathcal{O}$(100,
10~\rstar)), respectively. The corresponding methods base on the work by
\citet{WrightBarlow75}/\citet{PanagiaFelli75} (radio) and \citet{LW84a,
LW84b} (IR-excess). They have been applied, e.g., by \citet{ABC81} and
\citet{LL93} to derive OB-star radio mass-loss rates, and by \citet{LWW84}
to investigate the IR excess of $\zeta$ Pup by means of IRAS observations.
Since the involved processes are dominated by free-free and bound-free
transitions which scale with $\rho^2$, inferences from the
measurements are extremely sensitive to clumping in the wind. \\
\underline{3. \Ha\ emission,} usually modelled using NLTE atmosphere codes,
also depends on $\rho^2$, and is therefore also sensitive to clumping, but
samples the innermost portion of the wind ($\la{2}$\rstar). The idea to
exploit \Ha\ as a standard mass-loss indicator goes back to \citet{KC78},
and has been firstly applied by \citet{Leitherer88}, \citet{Drew90} and
\citet{LL93}. Further refinements have been provided by \citet{Pulsetal96}.

\section{Clumping -- some basic facts}

The present hypothesis states that clumping is a matter of {\it small-scale}
density inhomogeneities in the wind\footnote{in contrast to large-scale
inhomogeneities such as co-rotating interaction zones (e.g., \citealt{CO96}
and references therein).}, which redistribute the matter into clumps of
enhanced density embedded in a rarefied, almost void medium. The amount of
clumping is conveniently quantified by the so-called clumping factor, $\fcl
\ge 1$, which is a measure of the over-density inside the clumps (compared
to a smooth flow of identical average mass-loss rate).\footnote{An
alternative description bases on the volume-filling factor, $f_{\rm V} =
\fcl^{-1}$.} As already pointed out, diagnostics that are linearly dependent
on the density are insensitive to clumping, whilst those sensitive to
$\rho^2$ will tend to overestimate the mass-loss rate of a clumped wind, by
a factor $\sqrt{\fcl}$. For further details, see, e.g., \citet{ABC81, LW84b,
Schmutz95} and \citet{Pulsetal06}.

Until to date, the most plausible physical process responsible for small-scale 
structure formation in massive star winds is the so-called line-driven
instability, found already in the first time-dependent hydrodynamical
simulations of such winds \citep{OCR88}; for recent
investigations, see \citet{RO02, RO05} with respect to 1-D simulations and 
\citet{DO03, DO05} for first attempts to include 2-D processes.

Though predicted from the earliest hydro-models on (and even before, see
\citealt{LS70}), it took some while to incorporate clumping into the {\it
atmospheric models} of massive stars, firstly for Wolf-Rayet (WR)
atmospheres, in order to explain\\
(a) the strength of the observed electron
scattering wings ($\rho$-dependent) {\it in parallel} with the strength of the
underlying ($\rho^2$-dep.) emission lines \citep{Hillier91},\\ 
(b) the so-called momentum problem of WR winds (e.g., \citealt{Schmutz95}),\\
(c) the presence and variability of sub-structures in WR emission lines, 
by invoking supersonic turbulence leading to clumping factors of the order
of $\fcl \approx 9$ \citep{MoffatRobert94}.

\section{No clumping in OB-star winds?}

In contrast to the case of WR winds, the diagnostics of OB-star winds did not
render any necessity for (significant) clumping until recently. In
particular, two major arguments supported the view of a rather smooth,
almost unclumped flow:\\

\noindent
1. The investigations by \citet{LL93} and \citet{Pulsetal96} showed that
\Ha\ {\it and} radio mass-loss rates are similar for a large sample of stars
(but see also \citet{Drew90} who noted a discrepancy by a factor of two, the
former being larger). Since \Ha\ forms in the lower and the radio emission
in the outer wind, this would imply a similar degree of clumping in both
regions, which seemed to be improbable and is also in contradiction
to theoretical predictions.\\
2. The observed {\it wind-momentum rates} were found to be in rather good
agreement with theoretical predictions for this quantity, obtained from
different, independent investigations \citep{Vink00, Kudritzki02,
Pulsetal03, KK04}. A pure coincindence of this agreement seemed to be rather
unlikely.\\

\noindent Taken together, it was concluded that clumping effects in OB-star
winds should be negligible.

\section{Indications of (significant) clumping in OB-star winds}
\label{sigclump}

As stated in the introduction, recent evidence points to the notion that
this conclusion might be incorrect. In the following, we will summarize the
different indications for significant clumping in OB-star winds accumulated
so far. From the observational side, there is, to our knowledge, only one
{\it direct} evidence for clumping. From a temporal analysis of He{\sc ii}
4686, \citet[ in particular their Fig.~2]{Eversberg98} found "outward moving
inhomogeneities" in the wind of $\zeta$~Pup, from regions near the
photosphere out to 2 \rstar.

All other evidence is indirect, and the derived clumping factors cover a large
range.\\

\noindent
(a) From polarimetry and using simplified models, clumping factors of the
order of \fcl=2 are found (Davies, this volume).\\

\noindent
(b) The analysis of radio and submm data from $\epsilon$~Ori and $\zeta$~Pup
\citep{Blomme02, Blomme03} indicates a clear submm
excess in both cases, which can be explained by (enhanced) clumping in the
intermediate wind ($\approx 10 \rstar$) (see also Blomme, this volume).\\

\noindent
(c) NLTE model atmosphere analyses of UV spectra (partly incl. the optical
range) of various O-stars indicate clumping factors of the order of
$\fcl=10{\ldots}50$, but show that only few lines are suited to
discriminate between clumped and unclumped flows. From these analyses
(cf.~Table~\ref{tabfcl}), it was concluded that clumping, if present,
should begin rather close to the wind base, at a few tens of \kms. This
finding is in contrast to hydrodynamical simulations, which show that the
line-driven instability needs a certain time to grow and to become non-linear,
so that significant clumping is expected not before 1.2 to 1.3 \rstar. 

\begin{table}[h]
\caption{Evidence of clumping from UV diagnostics (NLTE).}
\footnotesize{
\tabcolsep2.0mm
\begin{tabular}{l l l l l}
\\
\hline
authors & objects & indicator & \fcl & comments \\
\\
\hline
Crowther et al.    & AV232(O7Iaf+) & P{\sc v}               & 10 & other lines barely affected\\ 
\, (2002)          & SMC           &                        &    & by clumping \\
Hillier et al.     & AV83(O7Iaf+)  & P{\sc v}/strong           & 10 & if clumping is important, \\ 
\, (2003)          & SMC           & UV photo-             &    & it must begin at relatively \\ 
		   &               & sph. lines                        &    & low velocities (30 km/s!)\\  
Bouret et al.      & SMC dwarfs    & O{\sc v}               & signi- &  \\
\, (2003)          &               &                        & ficant & \\
Bouret et al.      & HD\,190429A (O4If)  & P{\sc v}, O{\sc v}, N{\sc iv} & 25 & reduction of \mdot\ by factors  \\
\, (2005)	   & HD\,96715 (O4V((f)) &                               & 50 & of 5 and 7 \\  
                   &               &                        &    & clumping must start\\
                   &               &                        &    & at the wind base\\
\hline
\end{tabular}
}
\label{tabfcl}
\end{table}

\noindent
(d) From detailed investigations of the wind-momentum rates of a large
sample of O-stars, \citet{Pulsetal03, Markova04} and \citet{Repo04} found
that supergiants with \Ha\ in emission lie above the theoretical
wind-momentum luminosity relation (WLR), whereas the rest fits almost
perfectly.  Since the WLR should be independent of luminosity class (e.g.,
\citealt{Pulsetal96}), this discrepancy was interpreted in terms of
clumpy winds, with $\fcl \approx 5$, and mass-loss rates reduced by factors
between 2 and 3.\\ 

\noindent (e) A compelling, independent indication of clumping comes from
SEI analyses of the P-Cygni P\,{\sc v}\,$\lambda\lambda 1118, 1128$
resonance line doublet \citep[ see also Massa, this volume]{Massa03,
Fullerton06}, which has only become widely accessible since the launch of
{\it FUSE}. Because phosphorus has a low cosmic abundance, this doublet
never saturates in normal OB stars, providing useful estimates of not only
$\mdot q$ (cf. Sect.~\ref{diag}) but also \mdot\ itself, {\it when} P$^{4+}$
is the {\it dominant ion} -- as it is implied at least for mid-O star winds.
These mass-loss rates turned out to lie considerably below those inferred
from other diagnostics such as \Ha\ or radio emission. The most reasonable
way to reconcile these results is to invoke extreme clumping in the wind
(P{\sc v} as a resonance line remains insensitive to {\it direct} clumping
effects), indicating clumping factors of the order of $\fcl \approx 100$ (or
even more). Accordingly, the actual mass-loss rates should be {\it much} lower
than previously thought, by a factor of $\ga 10$. Further comments are given
in Sect.~\ref{pv}\\

\noindent
(f) Using similar methods, \citet{Prinja05} showed that the unsaturated P
Cygni lines in lower luminosity B supergiants give, again, a factor of 10
lower mass-loss rates than theoretically expected. \citet[ see also Crowter,
this volume]{Crowther06}, on the other side, found reasonable agreement
between observed and predicted mass-loss rates for early/mid BIa
supergiants, with similar (though unconstrained) clumping factors in the
lower and intermediate wind.

\section{A combined Ha/IR/mm/radio analysis}

Whereas most of the above investigations are concerned with a {\it global}
(i.e., radially almost constant) clumping factor or derive this quantity for
certain wind regions only, there is, of course, the additional question
regarding the radial stratification of \fcl. To this end, \citet{Pulsetal06}
recently performed a self-consistent analysis of \Ha, IR, mm and radio
fluxes, thus sampling the lower, intermediate and outer wind in parallel.
This study comprises a sample of 19 Galactic O-type supergiants and giants
with well known stellar parameters (from \citealt{Markova04, Repo04} and
\citealt{Mokiem05}), employing own measurements/archival data for \Ha, IR/mm
fluxes ({\it SCUBA}) and new {\it VLA} observations, including objects with
\Ha\ in absorption, i.e., low density winds. 

A major result of this investigation is that in {\it weaker} winds the
clumping factor is the same in the inner ($r < 2 \rstar$) and outermost
regions. However, for {\it stronger} winds, the clumping factor in the inner
wind is larger than in the outer one, by factors of 3 to 6. This finding
points to a physical difference in the clumping properties of weaker and
stronger winds (and may be related to the excitation mechanism of the
structure formation).  In terms of mass-loss rates then, we find
\mdot(radio) $\approx$ \mdot(\Ha) for weak winds with \Ha\ in absorption,
whereas for all stars with \Ha\ in emission we obtain \mdot(radio) $\approx$
0.4{\ldots}0.5 \mdot(\Ha), consistent with the arguments given by Markova et
al./Repolust et al. and the earlier findings by \citet{Drew90}.

Unfortunately, this analysis is hampered by one severe restriction. Since
{\it all} diagnostics employed have a $\rho^2$ dependence, only {\it
relative} clumping factors could be derived, normalized to the values in the
outermost, radio-emitting region. In other words, \mdot({\sc real}) $\le$
\mdot(radio), since until now the clumping in the radio emitting region is
still unknown. Only if \fcl(radio) were unity, we would have \mdot({\sc real}) =
\mdot(radio). Thus, the issue of absolute values for \mdot\ still remains
unresolved. 

\begin{figure}
\center{\includegraphics[width=11cm]{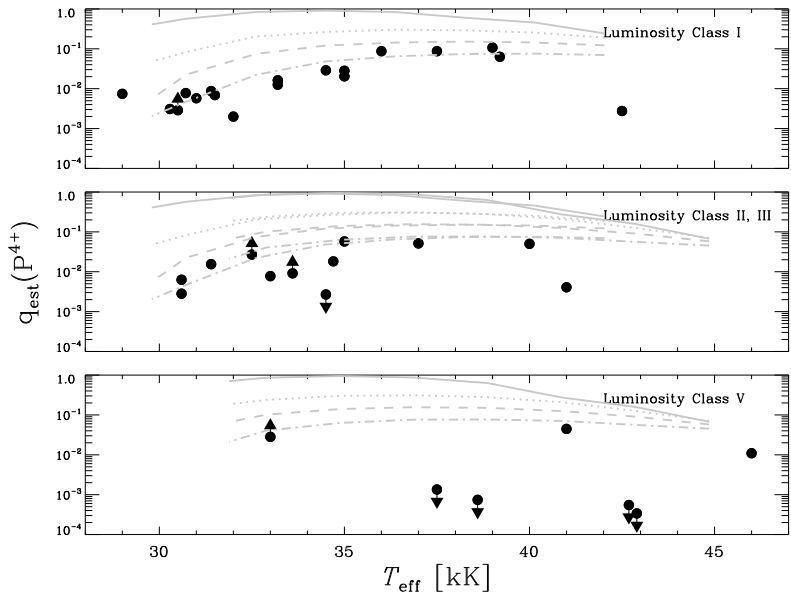}}
\caption{Derived estimates for $q_{\rm est}$ (Eq.~\ref{qesteq}) of P$^{4+}$, as a function of
\teff\ and luminosity class (for details, see \citealt{Fullerton06}).
Overplotted are predicted values for this quantity, as obtained from own
calculations. Bold: unclumped models; other curves: clumped models, with \fcl=
9, 36, 144 (dotted, dashed, dashed-dotted), respectively.} 
\label{qest}
\end{figure}

\section{The PV problem}
\label{pv}

The major result of the investigation by \citet{Fullerton06} (see
Sect.~\ref{sigclump}) is displayed in Fig.~\ref{qest}. In this figure, they
have plotted, as a function of \teff\ and luminosity class, the quantity
\begin{equation}
q_{\rm est} = \frac{\langle \mdot q \rangle_{\rm obs}}{\mdot_{\rm Ha/radio}}
\quad \rightarrow \frac{\langle \mdot q \rangle_{\rm obs}}{\mdot \sqrt{\fcl}}
\quad \rightarrow \frac{\langle q \rangle}{\sqrt{\fcl}},
\label{qesteq}
\end{equation}
for P$^{4+}$, where angle brackets denote spatial averages.  The quantity in
the nominator has been measured from the P{\sc v} lines, and the
corresponding \Ha/radio mass-loss rates have been taken from the literature.
If the winds were unclumped, $q_{\rm est}$ would correspond to the average
ionization fraction of P$^{4+}$, whereas for clumped winds this quantity is
modified by $\fcl^{-0.5}$. Fullerton et al. now argue that P$^{4+}$ should
be a dominant ion at mid O-type, and in this case the derived clumping
factor would be $\fcl = \mathcal{O}(100)$ at \teff $\approx$ 40,000~K.
Additionally, however, they report on test calculations performed with
{\it unclumped} models which show that, on the contrary, P$^{4+}$ should
become a dominant ion only below O7. If this were true, Fig.~\ref{qest}
would imply $\fcl \approx 10,000$ in this temperature regime!

\begin{figure}
\center{\includegraphics[angle=90, width=11cm]{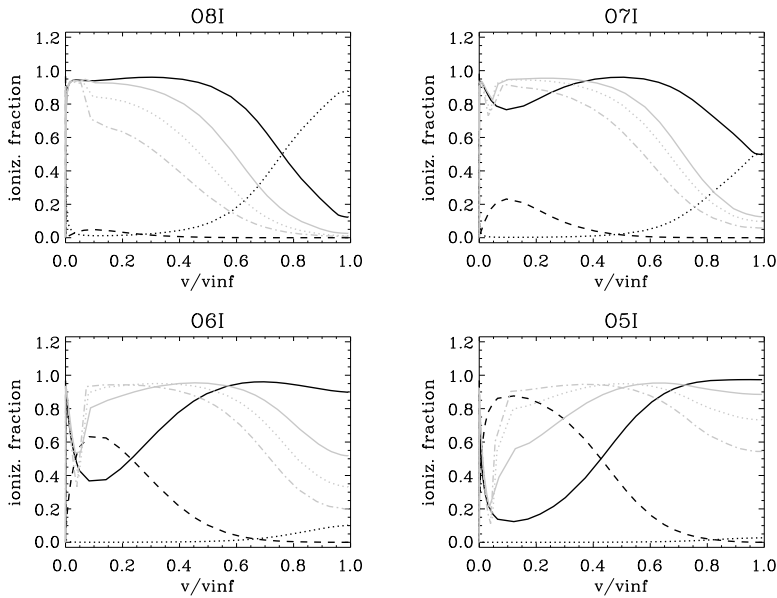}}
\caption{Ionization fractions of phosphorus, as a function of velocity and
spectral type, for supergiants. Black: Unclumped models (bold - P{\sc v},
dotted - P{\sc iv}, dashed - P{\sc vi}). Grey: Ionisation fractions of 
P{\sc v}, for clumped models with \fcl = 9, 36, 144 (bold, dotted, dashed),
respectively. } 
\label{pvifrac}
\end{figure}

Since this is VERY unlikely, we have investigated the influence of clumping
on the ionization fraction of P$^{4+}$. Due to the enhanced density inside
the clumps, stronger recombination is expected (this is the {\it indirect}
effect of clumping), which should change the picture (see also
\citealt{Bouret05}). By means of our model atmosphere code FASTWIND
\citep{Puls05} and using the phosphorus model atom provided by
\citet{Pauldrach01}, we have calculated a sequence of clumped models with
$\mdot \sqrt{\fcl}$ = const, i.e., models which have identical
$\rho^2$-diagnostics mass-loss rates. As shown in Fig.~\ref{pvifrac},
increased clumping indeed shifts P$^{4+}$ as a dominant ion towards higher
\teff: for unclumped models, it is dominant at O8/7, whereas for the largest
clumping factors it dominates at O5.

Using these models then, the ``observed'' run of $q_{\rm est}$ can be
reproduced with highly clumped models (\fcl = 144, see Fig.~\ref{qest}), for
almost all luminosity classes and except for the very hottest temperatures
(see below). Thus, our calculations confirm the hypothesis stated by
Fullerton et al., i.e., clumping {\it is} possible to explain the observations,
and mass-loss rates might indeed be lower by factors of 10 or even more.

\section{Implications}

As we have seen, there seems to be a (physical) difference between thinner
and thicker winds. For thinner winds, there is a similar degree of clumping
in the lower and outer wind, whereas for thicker winds clumping is stronger
in the lower part. The actual mass-loss rates depend on the clumping in outer
wind, which is still an unsolved issue.

\noindent
{\it If} the outer winds were unclumped, our results would be consistent
with theoretical WLRs. In this case then, one would meet a severe dilemma
with the results from the F(UV), which might hopefully be explained by
additional effects from X-rays emitted due to clump-clump collisions
\citep{Feld97, Pauldrach01}. X-rays might also help to solve the problem
encountered for P{\sc v} at highest \teff.

\noindent
{\it If}, on the other hand, the (F)UV values were correct, the outer wind
must be significantly clumped, and the present match of "observed" and
predicted WLR would indeed by only coincidental. This scenario would imply a
number of severe problems, not only for radiation driven wind theory, but,
most importantly, concerning the stellar evolution in the upper HRD and
related topics. A possible way out of the latter problem has been suggested
by N. Smith (this volume).

%\section{}         %%% Top level section head (remove "%" symbol)
%\subsection{}      %%% Second level section head (remove "%" symbol)
%\subsubsection{}   %%% Lowest level section head (remove "%" symbol)
%\section*{}	    %%% Unnumbered top level section head (remove "%" symbol)
%\subsection*{}     %%% Unnumbered second level section head (remove "%" symbol)

\noindent
\acknowledgements 
Part of this work has been supported by the NSF of the Bulgarian
Ministry of Education and Science (No. 1407/2004).

%%% THE BIBLIOGRAPHY
%%%
%%% CONSULT SECTION 3 OF "INSTRUCTIONS FOR AUTHORS" FOR HOW TO USE NATBIB.
%%% AUTHORS ARE ENCOURAGED TO USE EITHER THE "THEBIBLIOGRAPY" ENVIRONMENT
%%% BY UNCOMMENTING (DELETING THE "%" SYMBOL) THE COMMANDS BELOW, OR BY
%%% USING THE BIBTEX ENVIRONMENT. TO FIND OUT WHICH IS APPLICABLE TO YOUR
%%% CONTRIBUTION, CONSULT THE VOLUME EDITORS FOR YOUR PROCEEDINGS.
%%%

\end{document}